\documentclass[preprint,showpacs,preprintnumbers,amsmath,amssymb]{revtex4}

\usepackage{graphicx}
\usepackage{dcolumn}
\usepackage{bm}
\newcommand {\bsub} {\begin{subequations}}
\newcommand {\esub} {\end{subequations}}
\newcommand {\bea} {\begin{eqnarray}}
\newcommand {\eea} {\end{eqnarray}}
\def\ie{{\it i.e.}}
\def\eg{{\it e.g.}}

\newcommand{\lsim}{ \mathop{}_{\textstyle \sim}^{\textstyle <} } 
 
\begin{document}

\preprint{OCHA-PP-236}

\title{
Neutron electric dipole moment and flavor changing interactions 
\\
in supersymmetric theories 
}

\author{Gi-Chol Cho}
\affiliation{%
Department of Physics, Ochanomizu University,\\
Tokyo, 112-8610, Japan
}%
\author{Naoyuki Haba}
\affiliation{%
Institute of Theoretical Physics, University of Tokushima, \\
Tokushima 770-8502, Japan
}%
\author{Minako Honda}
\affiliation{%
Graduate School of Humanities and Sciences, \\
Ochanomizu University, Tokyo, 112-8610, Japan
}%

\begin{abstract}
Supersymmetric contributions to the neutron electric dipole moment (EDM) 
are studied taking account of the flavor changing interactions. 
We found that the gluino contribution is sensitive to the flavor 
changing interaction. 
Enhancement of neutron EDM via flavor mixing effects is possible when 
the squark mass difference between the different generations is sizable. 
As an example, the results of the SUSY SU(5) GUT with right-handed neutrinos 
are briefly discussed. 
\end{abstract}

\maketitle
Supersymmetry (SUSY) provides an elegant solution to the gauge hierarchy 
problem. 
It is, however, known that SUSY must be broken softly at the Fermi scale,  
$G_F^{-1/2}$, since there has been no evidence of a boson whose mass 
is exactly same with one of known fermions and vice versa. 
Introducing soft SUSY breaking terms can increase the mass of SUSY particles, 
but its scale must be $O({\rm TeV})$ to make the electroweak scale 
stable against radiative corrections. 
%

%
One of the serious problems of softly broken supersymmetric Standard 
Model (SUSY-SM) is a possibility of $CP$ violation beyond the 
Kobayashi-Maskawa mechanism. 
In general, a softly broken SUSY Lagrangian contains a lot of complex 
parameters. 
However, in a certain class of SUSY breaking scenarios, 
most of them could be removed by field re-definition. 
For example, in the gravity mediated SUSY breaking scenario which leads
to the universal scalar mass, trilinear coupling and gaugino mass, 
it is known that there are two complex phases as physical degree of 
freedom, in addition to the Kobayashi-Maskawa 
phase~\cite{Dugan:1984qf}. 
Conventionally two complex phases are chosen as those of the scalar 
trilinear coupling $A_f$ and the higgsino mass $\mu$: 
\begin{eqnarray}
A_f = |A_f| e^{i\alpha_f}, ~~~\mu = |\mu| e^{i\phi}, 
\label{cpphase}
\end{eqnarray}
where $f$ denotes the flavor of (s)quarks or (s)leptons. 
Stringent constraints on the parameters $\alpha_f$ and $\phi$ 
are given by the upper bound of electric dipole moment (EDM) 
of neutron~\cite{Harris:1999jx}: 
\begin{eqnarray}
|d_n| = 6.3 \times 10^{-26}{e\cdot {\rm cm}}. 
\label{edm}
\end{eqnarray}
The supersymmetric contributions to the neutron EDM consist of the 
following 1-loop diagrams: (i)squark-gluino, (ii) squark-chargino 
and (iii) squark-neutralino exchanges. 
It has been pointed out that a natural size $(\sim O(1))$ of $\alpha_f$ 
and $\phi$ is strongly disfavored from the neutron EDM measurements, 
unless the squarks are heavier than a few TeV~\cite{Kizukuri:1992nj}. 
However, so far constraints on squark masses from the neutron EDM have 
been studied without paying a special attention to the flavor off-diagonal
interactions between quarks and squarks, though it is allowed in general. 
For example, some class of supersymmetric grand unified theories 
(SUSY GUT) predict that a large mixing of neutrinos leads to a sizable 
generation mixing of right-handed down-squarks, which could be observed as 
an enhancement of flavor changing interactions between quarks and 
(right-handed) down-squarks~\cite{Akama:2001em}. 
Also, recently reported deviation of $CP$ asymmetry in 
$b \to s q \bar{q}$ from the SM expectation 
at B-factories~\cite{ichep_belle,ichep_babar} 
motivates us to study a possibility of $CP$ violation in SUSY flavor 
changing interactions. 
%

%
In this paper we investigate effects of the flavor changing interactions
between quarks and squarks to the neutron EDM model independently, when 
the SUSY $CP$ phases (\ref{cpphase}) are not suppressed. 
Our main interest is that if the lower mass bound on the SUSY particles 
from the neutron EDM is altered significantly due to the 
flavor changing interactions. 
The works in refs.~\cite{Endo:2004xt} have been done in a similar 
direction with our study. 
They examined the squark flavor mixing based on some class of SUSY-GUT, 
emphasizing the anomaly in $CP$ asymmetry of $B_d \to \phi K_S$. 
Contrary to the previous studies, we analyze the flavor mixing effect 
phenomenologically, allowing somewhat large flavor changing
interactions. 
We find that the gluino mediated diagram is sensitive to the flavor 
mixing effect while the chargino and neutralino diagrams are not. 
However the flavor mixing effects are suppressed when the mass 
difference of squarks between different generations is small enough. 
%

%
The squark mass matrix $M_{\widetilde{q}}^2~(q=u,d)$ in the generation space 
is given by 
\begin{eqnarray}
M_{\widetilde{q}}^2 &=& \left(
\begin{array}{cc}
(M_{\widetilde{q}}^2)_{LL} & (M_{\widetilde{q}}^2)_{LR} \\
(M_{\widetilde{q}}^2)_{RL} & (M_{\widetilde{q}}^2)_{RR} 
\end{array}
\right), 
\end{eqnarray}
where $(M_{\widetilde{q}}^2)_{\alpha\beta}~(\alpha,\beta=L,R)$ 
is a $3\times 3$ matrix. 
The generation mixing of $LL$ and $RR$ parts can be removed 
by transforming a basis of squark $\widetilde{q}_\alpha^0$ as 
\begin{eqnarray}
\widetilde{q}^0_\alpha &=& \widetilde{U}^q_\alpha \widetilde{q}_\alpha. 
\label{unitary}
\end{eqnarray}
The off-diagonal elements, 
$(M_{\widetilde{q}}^2)_{LR}$ or $(M_{\widetilde{q}}^2)_{RL}
(=(M_{\widetilde{q}}^2)_{LR}^{\dagger})$, are then given 
as follows: 
\begin{eqnarray}
(\widetilde{U}^q_L)^\dagger (M_{\widetilde{q}}^2)_{LR} 
\widetilde{U}^q_R 
&=& 
(\widetilde{U}^q_L)^\dagger U^q_L
M_q^{\rm diag} A_{{\rm eff},q}^*  
(U^q_L)^\dagger \widetilde{U}^q_R, 
\\
A_{{\rm eff},q} &=& A_q^* - \mu T_q, 
\label{lrmixing}
\end{eqnarray}
where $T_q$ is given as $T_u =\cot\beta$  and 
$T_d =\tan\beta$. 
$\tan\beta$ is defined as a ratio of $v_u$ and $v_d$ which 
are vacuum expectation values of two Higgs doublets, $H_u$ and $H_d$,
respectively. 
The unitary matrices $U^q_L$ and $U^q_R$ diagonalizes the quark mass 
matrix as
\begin{eqnarray}
(U^q_L)^\dagger M_q U^q_R = {\rm diag}(m_{q1},m_{q2},m_{q3}), 
\end{eqnarray}
where 
$(m_{q1},m_{q2},m_{q3})=(m_u,m_c,m_t)$ for $q=u$ and $(m_d,m_s,m_b)$ for 
$q=d$. 
The Cabibbo-Kobayashi-Maskawa (CKM) matrix is defined as
\begin{eqnarray}
V_{\rm CKM} \equiv U_L^{u\dagger} U_L^d. 
\label{ckm}
\end{eqnarray}
Since the off-diagonal parts of the mass matrix, eq.~(\ref{lrmixing}), 
is suppressed by a quark mass comparing with the diagonal parts 
when the soft SUSY breaking terms are around $O({\rm TeV})$. 
In general, the unitary matrices $\widetilde{U}^q_\alpha$ and 
$U^q_\alpha$ are not coincide. 
This is a source of flavor changing interactions between quarks and squarks. 
%

%
Next let us see the SUSY interactions relevant to the neutron EDM. 
There are two types of interactions -- the gaugino-quark-squark 
interaction and the higgsino-quark-squark interaction.  
An example of the former is the interactions of gluino $\widetilde{g}$
to quarks and squarks. After removing the generation mixing, the 
interaction Lagrangian is given by 
\begin{eqnarray}
{\cal L} &=& 
-\sqrt{2}g_s \overline{\widetilde{g}^a} T^a 
\left\{
\widetilde{u}_L^* V_{LL}^u u_L  - \widetilde{u}_R^* V_{RR}^u u_R 
\right.
\nonumber \\
&&~~~
\left.
+
\widetilde{d}_L^* V_{LL}^d d_L  - \widetilde{d}_R^* V_{RR}^d d_R 
\right\} 
+{\rm h.c.}, 
\label{lagrangian2}
\end{eqnarray}
where $g_s$ and $T^a$ are the SU(3)$_C$ gauge coupling and the color
matrix, respectively. 
The unitary matrix $V_{\alpha\beta}^q$ is defined as 
\begin{eqnarray}
V_{\alpha\beta}^q \equiv 
\widetilde{U}_\alpha^{q\dagger} 
U_\beta^q. 
\label{unitary3}
\end{eqnarray}
Eq.~(\ref{lagrangian2}) tells us that the gluino diagram can
contribute to the neutron EDM only through the left-right mixing 
of squarks in the 1-loop propagator. 
The left-right mixing is induced by the off-diagonal element of squark
mass matrix (\ref{lrmixing}), which is proportional to the quark mass. 
Thus the gluino diagrams with flavor changing interactions 
could be enhanced by the left-right mixing of squark propagators beyond 
the 1st generation. 
%

%
The chargino or neutralino can couple to the quarks and squarks through 
the Yukawa interactions, in addition to the gaugino-quark-squark 
interaction. 
The interaction of the right-handed down-quark to the up-squark and the
charged higgsino is given as 
\begin{eqnarray}
{\cal L} &=& -(f^d)^*_{ij} \widetilde{u}^{0*}_{Li} 
\overline{\widetilde{H}^-_{dL}} d_{Rj}^0 + {\rm h.c.}, 
\label{higgsino}
\end{eqnarray}
where $i,j$ denote the generation index. 
Note that the Yukawa matrix $f_d$ is related to the quark mass matrix 
as $M_d = -(f^d)^* v_d$. 
Using the unitary matrices $\widetilde{U}^q_\alpha$ and $U^q_\alpha$, 
the interaction (\ref{higgsino}) is written as 
\begin{eqnarray}
{\cal L} &=& 
(\widetilde{u}^*_L)_i
\overline{\widetilde{H}^-_{dL}} 
\left\{\widetilde{U}^{u\dagger}_L U_L^d 
\frac{g\, {\rm diag.}(m_d,m_s,m_b)}{\sqrt{2}m_W\cos\beta}
\right\}_{ij} 
(d_R)_j \nonumber \\ 
&&~~~~
+ {\rm h.c.}. 
\label{higgsino2}
\end{eqnarray}
In the case of $d$-quark EDM ($j=1$), therefore, the interaction is
proportional to $m_d$. 
The complete set of the interactions will be shown explicitly in our 
subsequent paper~\cite{CHH}. 
%

\begin{figure}[ht]
\includegraphics[width=5cm,clip]{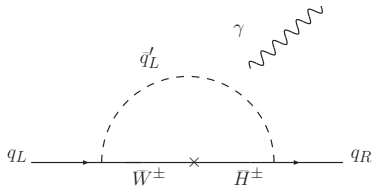}
\caption{
Feynman diagram of chargino contributions to the neutron EDM.
}
\label{fig04}
\end{figure}

%
It is well known that, in the flavor diagonal case, the GUT relation 
for the gaugino masses, 
$M_3:M_2:M_1 = \alpha_3:\alpha_2:\frac{5}{3}\alpha_Y$, 
makes the chargino much lighter than the gluino so that the chargino
diagrams give the dominant contributions~\cite{Kizukuri:1992nj}. 
The neutralino contribution is always small. 
We will see that the gluino contribution is quite sensitive to the 
flavor changing interaction while the chargino contribution is not. 
We show in Fig.~\ref{fig04} the 1-loop diagram which gives the leading 
contribution in the chargino exchanging diagrams. 
This is comprised of 
$\widetilde{W}^\pm$-$q_L$-$\widetilde{q}'_L$ and 
$\widetilde{H}^\pm$-$q_R$-$\widetilde{q}'_L$ vertices, the former is the
SU(2)$_L$ version of (\ref{lagrangian2}) and the latter is
(\ref{higgsino2}). 
As is already mentioned, the interaction (\ref{higgsino2}) is always
proportional to the 1st generation quark mass while (\ref{lagrangian2}) 
is not enhanced by the flavor off-diagonal components because of the 
gauge interaction. 
This is why the chargino contribution is insensitive to the flavor changing
interactions. 
In this paper, therefore, we consider the flavor changing gluino 
interaction only, where the flavor mixing is parametrized by 
(\ref{unitary3}) with $\alpha=\beta$. 
%

%
Throughout this paper, we will perform our numerical study to the 
two generation case. 
Then the flavor mixing matrix (\ref{unitary3}) is parametrized as
\begin{eqnarray}
V^q=\left(
\begin{array}{cc}
\cos\theta_q & \sin\theta_q\\
-\sin\theta_q & \cos\theta_q
\end{array}
\right). 
\end{eqnarray}
The extension to the three generation case is straightforward, 
and it will be shown in our next paper~\cite{CHH}. 
%

%
\begin{figure}[ht]
\includegraphics[width=6cm,clip]{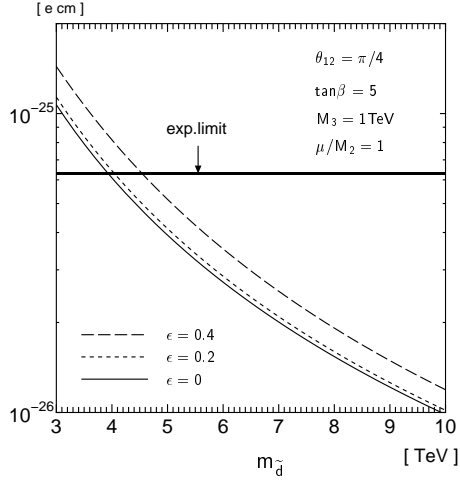}
\caption{
Supersymmetric contributions to the neutron EDM for $\tan\beta=5$. 
The flavor mixing angle is taken as $\pi/4$. 
The $CP$ violating phases are taken as $\alpha_f= \phi=\pi/4$. 
The gluino mass $M_3$ is $1{\rm TeV}$. 
The chargino and neutralino contributions are obtained using 
the GUT relation and $\mu/M_2=1$. 
}
\label{fig01}
\end{figure}
Next we show our numerical result for SUSY contributions to the neutron
EDM taking account of the flavor off-diagonal interactions. 
It is already discussed that only the gluino contribution can give 
non-negligible contributions to the neutron EDM via flavor changing 
interactions. 
As is mentioned earlier, the transition between the left- and 
right-handed squarks are suppressed by the small quark mass as compared 
to the chirality preserved case. 
In our calculation, therefore, the effects of left-right mixing of squarks 
are taken into account using the mass insertion approximation, using 
(\ref{lrmixing}) as an expansion parameter. 
%

%
In the following, we adopt the GUT relation for gaugino masses 
to reduce the number of parameters. 
The left- and right-handed squark masses are taken to be equal. 
We also assume all the squark flavor mixing angles 
between the 1st and 2nd generations are common, 
\ie, $\theta^u_{\alpha\alpha}=\theta^d_{\alpha\alpha} 
\equiv \theta_{12}$ for simplicity. 
We suppose that the scalar trilinear term $A_f$ is universal 
in the 
generation space and take as $A_f=m_{\widetilde{Q}}/3$. 
In the numerical study, we introduce a parameter $\epsilon^d$, 
\begin{eqnarray}
\epsilon^d \equiv 
\frac{m^2_{\widetilde{d}}-m^2_{\widetilde{s}}}
{m^2_{\widetilde{d}}}
\label{splitting}
\end{eqnarray}
in order to take account of the mass difference 
between the 1st and 2nd generation down squarks. 
Another parameter $\epsilon^u$ is also introduced in the up 
squark sector. 
We will, however, take the same values for the up and down 
squarks for each generation, so that the parameters $\epsilon^u$ 
and $\epsilon^d$ are denoted by $\epsilon$ in the following. 
%

%
Fig.~\ref{fig01} shows the neutron EDM as a function of the 1st
generation squark mass, taking account of the flavor changing 
interactions. 
The gluino mass is fixed at 1 TeV and the flavor mixing angle 
is taken as $\theta_{12}=\pi/4$. 
The parameter $\epsilon$ is taken as 
$0,0.2$ and $0.4$, as shown explicitly in the figure. 
The $CP$ violating phases $\alpha_f, \phi$ in (\ref{cpphase}) 
are fixed at $\pi/4$, and $\tan\beta$ is 5. 
The horizontal line denotes the upper bound of neutron EDM (\ref{edm}). 
The chargino and neutralino contributions are obtained using the GUT 
relation on the gaugino masses and $\mu/M_2=1$. 
We can see in the figure that the flavor off-diagonal interactions 
increase the neutron EDM as the violation of squark mass degeneracy 
increases. 
Note that the enhancement of the neutron EDM comes from the gluino 
diagram. 
It is interesting that, when the squark mass universality is sizably 
violated ($\epsilon=0.4$), the flavor mixing effect somewhat increases 
the squark lower mass bound. 
It is known that the squark mass splitting between the first two 
generations is strongly constrained from the $K^0$-$\overline{K^0}$ 
mixing as 
$\sin^2\theta \epsilon^2 
\left(\frac{30{\rm TeV}}{m_{\widetilde{d}}}\right)^2
\lsim 1$~\cite{Hisano:2000wy}.
Then, the mass splitting 
$\epsilon=0.4$ with about 5 TeV squark mass is excluded from the
$K^0$-$\overline{K^0}$ mixing, when the flavor mixing angle is $\pi/4$. 
The reason why the flavor mixing effect is marginal in the universal
limit of squark masses is as follows. 
In the gluino diagram, the flavor transition can occur at the 
quark-squark-gluino vertices and at the left-right mixing of 
squarks in the loop. 
Owing to the unitarity of the flavor mixing matrix, the diagrams 
proportional to the 2nd or the 3rd generation quark mass through the 
left-right mixing vanish when there is no mass difference of 
squarks among the different generation. 
%

%
\begin{figure}[ht]
\includegraphics[width=6cm,clip]{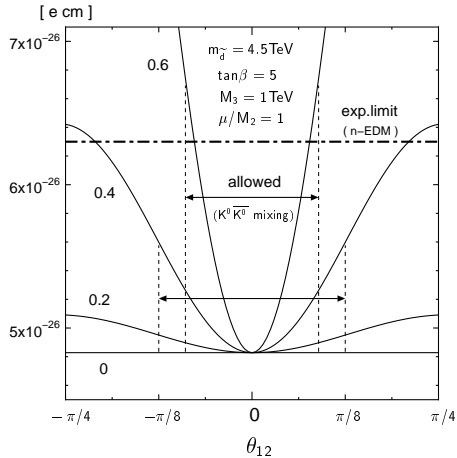}
\caption{
Supersymmetric contributions to the neutron EDM as functions of 
the mixing angle for $\tan\beta=5$. 
The squark mass is fixed at $4.5{\rm TeV}$. 
}
\label{fig02}
\end{figure}
The flavor mixing angle dependence of the SUSY contributions to the  
neutron EDM is shown in Fig.~\ref{fig02} for $\tan\beta=5$. 
The 1st generation squark masses are fixed at $4.5{\rm TeV}$. 
In the figure, each line represents the violation of squark mass
universality from $\epsilon=0$ to $0.6$. 
This figure tells us that there is a strong correlation between 
the flavor mixing effect and the violation of the squark mass 
universality. 
It is easy to see that the sensitivity of flavor mixing angle becomes
strong as the mass difference of squark masses in the generation space
is large. 
In the figure, constraints on the mixing angle from the 
$K^0$-$\overline{K^0}$ mixing are shown explicitly for each value 
of $\epsilon$. 
\begin{figure}[ht]
\includegraphics[width=6cm,clip]{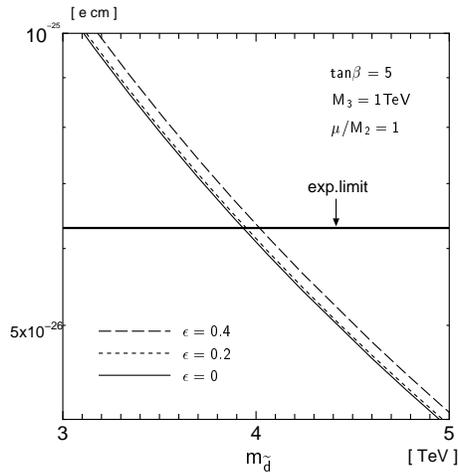}
\caption{
The neutron EDM in the SUSY SU(5) + $\nu_R$. 
}
\label{fig03}
\end{figure}
%

%
We have so far studied contributions due to the flavor changing 
interaction between quarks and squarks to the neutron EDM in general 
framework. 
The SUSY contribution is much enhanced by the large flavor mixing 
in the first two generations when the squark mass universality is 
violated significantly. 
However most of such parameter region is already disfavored from 
the $K^0$-$\overline{K^0}$ mixing. 
The similar enhancement of neutron EDM is possible for the flavor mixing 
between the 1st and 3rd generations, which is less constrained from the 
FCNC processes~\cite{CHH}. 
Then we might find severe constraints on so called ``decoupling''
solution to the SUSY FCNC problem (see, \cite{Hisano:2000wy} and 
references therein). 
%

%
The magnitude of the mixing angle is predicted when a specific SUSY
breaking scenario is chosen. 
Here we would like to discuss consequences of the SUSY SU(5) GUT with 
the right-handed neutrinos, which is one of the well-motivated GUT models. 
In this model, the flavor off-diagonal interactions are induced through
the radiative corrections to the soft SUSY breaking masses for the
scalar fields. 
If the up-quark and the Dirac neutrino Yukawa couplings are sufficiently
larger than the down-quark Yukawa couplings, we find that the 
$d_L$-$\widetilde{d}_L$ transitions are parametrized by the CKM matrix
while $d_R$-$\widetilde{d}_R$ transitions are proportional to the MNS 
matrix in a good approximation~\cite{CHH}. 
On the other hand, $u_L$-$\widetilde{u}_L$ and 
$u_R$-$\widetilde{u}_R$ transitions are flavor diagonal. 
The contributions to the neutron EDM in this framework are shown in 
Fig.~\ref{fig03} for $\tan\beta=5$. 
It can be seen in the figure that there is an enhancement via the flavor 
changing effects but its magnitude is small as compared to 
Fig.~\ref{fig01}. 
This is because that, in this framework, the flavor changing interactions 
are allowed only in the $d$-quark EDM, and those in the gluino diagram is  
parametrized by the CKM and MNS matrix elements. 
The enhancement of the gluino contribution, therefore, is much 
suppressed as compared to the cases in Figs.~\ref{fig01} and \ref{fig02}.  
%

%
The GUT relation for the gaugino masses makes the
gluino mass heavier, in order to satisfy the lower mass bound on the chargino, 
$m_{\widetilde{\chi}^-_1}>104{\rm GeV}$~\cite{lep_chargino}. 
If, however, one allows more lighter gluino, \eg, as light as its lower
mass bound from collider search experiments, 
$M_3 > 195{\rm GeV}$~\cite{Affolder:2001tc}, 
the gluino diagram is overwhelmingly dominant in the SUSY contributions
to the neutron EDM and makes the squark lower mass bound much severe. 

We add a comment on the electron EDM. The lack of the gluino
contribution makes the enhancement marginal. We will discuss this issue 
in ref.~\cite{CHH} in more detail.

\begin{acknowledgments}
This work is supported in part by the Grant-in-Aid for Science 
Research, Ministry of Education, Science and Culture, Japan,  
No.13740149 (G.C.C.) and No.16540258, 16028214, 14740164 (N.H.). 
\end{acknowledgments}

\end{document}